\documentstyle[12pt]{article}
\newcommand{\nc}{\newcommand}
\nc{\jv}{\frac{J}{4}}
\nc{\neel}{Ne\'{e}l}
\nc{\qq}{\P}
\nc{\rng}{\rangle}
\nc{\lng}{\langle}
\nc{\rcite}{ref.\ \cite}
\nc{\ba}{\begin{array}}
\nc{\ea}{\end{array}}
\nc{\Id}{I}
\nc{\Nul}{{\bf 0}}
\nc{\reals}{{\sf I}\kern-.12em{\sf R}}
\nc{\compl}{{\sf I}\kern-.48em{\sf C}}
\nc{\Z}{{\sf Z}}
\nc{\lb}{\left(}
\nc{\rb}{\right)}
\nc{\qrt}{\frac{1}{4}}

\newcommand{\bt}{\beta}
\newcommand{\gm}{\gamma}
\newcommand{\dl}{\delta}

\newcommand{\noi}{\noindent}
\newcommand{\half}{\frac{1}{2}}
\newcommand{\hhalf}{{\textstyle \frac{1}{2}}}

\newcommand{\rf}[1]{(\ref{#1})}

\newcommand{\ra}{\rightarrow}
\newcommand{\be}{\begin{equation}}
\newcommand{\ee}{\end{equation}}
\newcommand{\bea}{\begin{eqnarray}}
\newcommand{\eea}{\end{eqnarray}}
\nc{\nn}{\nonumber}
\nc{\thpm}{'t~Hooft-Polyakov monopole}
\nc{\eqrf}{eq.\ \rf}
\nc{\site}{x}
\nc{\shat}{\hat{S}}
\nc{\Tr}{\mbox{Tr}\;}
\nc{\defeq}{\stackrel{{\rm def}}{=}}
\nc{\calo}{{\cal O}}

\newcommand{\plb}[1]{Phys.~Lett.~#1B\ }

\newcommand{\npb}[1]{Nucl.~Phys.~B#1\ }
\newcommand{\nproc}[1]{Nucl.\ Phys.~B (Proc.~Suppl.)~#1\ }

\nc{\ACK}{\vspace{2ex} \noi {\bf Acknowledgement\\[1ex]}}
\begin{document}
\hspace*{\fill} OUTP-92-16P \\
\hspace*{\fill} hep-lat/9208013 \\

\vspace*{2.5cm}

 \renewcommand{\thefootnote}{\fnsymbol{footnote}}

\begin{center}
{\LARGE\bf
     Gauge invariant extremization\\ on the lattice\\[3 cm]
}
\normalsize

     A.J.\ van der Sijs\footnote{Supported by
     SERC grant GR/H01243.}\\[5mm]
     Theoretical Physics, University of Oxford\\
     1 Keble Road, Oxford\ \ OX1 3NP, United Kingdom\\
{\small    [e-mail: {\tt vdsijs@dionysos.thphys.ox.ac.uk}]\\[1.0cm]}

\normalsize
\vspace{5\baselineskip}

{\bf Abstract}\\
\end{center}

\noi
Recently, a method was proposed and tested to
find saddle points of the action in simulations
of non-abelian lattice gauge theory.
The idea, called `extremization', is to minimize
$\int(\dl S/\dl A_\mu)^2$.
The method was implemented in an explicitly gauge variant way,
however, and gauge dependence showed up in the results.

Here we show how extremization can be formulated in a way
that preserves gauge invariance on the lattice. The method applies to
any gauge group and any lattice action. The procedure is
worked out in detail for
the standard plaquette action with gauge groups U(1) and SU(N).

\thispagestyle{empty}

\newpage

\setcounter{footnote}{0}
\renewcommand{\thefootnote}{\arabic{footnote}}

\section{Introduction}

One of the approaches to understand the non-perturbative
nature of non-abelian gauge theories has been to look for topological objects
like monopoles and instantons in lattice Monte Carlo simulations.
They are local minima of the action and are thought to give important
contributions to the path integral.
An important question is whether the presence of such objects is
related to the string tension.
The problem one faces here is that short distance fluctuations obscure
the presence of such topological objects
in lattice gauge field configurations.
Monopoles and instantons are basically semiclassical objects,
characterized by their long distance structure.

Several techniques have been developed to extract the relevant long
distance information from a lattice gauge field configuration.
One of them is the cooling method \cite{cooling},
which minimizes the action density of
a particular configuration
locally in an iterative way,
directing it towards a local minimum in
configuration space.
Apart from local minima, however, saddle points are also solutions
to the field equations.
The sphaleron is a popular example.
It is therefore useful to consider saddle points in lattice field
configurations as well.

Recently, Duncan and Mawhinney \cite{dunmaw1} have discussed
a method to find saddle point solutions of Yang-Mills
theory (or other (lattice) field theories).
They applied their method, which they called `extremization',
to three-dimensional
SU(2) gauge theory and found that configurations after extremization showed
lumps in the action density that were highly correlated to lumps found
after cooling the same configurations.
It was found that the string tension survives moderate extremization but
ultimately disappears.

The basic idea of extremization is to minimize the squared gradient of the
action, referred to as `extremization action',
\be
\shat = \sum_{\site,\mu,a} \lb \frac{\dl
S}{\dl A_{\mu}^a (\site)} \rb^2  \; , \label{101}
\ee
instead of lowering the action itself as in the cooling method.
In ref.~\cite{dunmaw1} this was implemented for the plaquette
action of lattice gauge theory. A conjugate gradient method in terms of
the gauge fields $A_{\mu}^a(\site)$ was used to minimize $\shat$ and
Fourier acceleration was used to speed up the minimization process.
The authors stated that $\shat$ is a gauge
{\em variant\/} function, so that lattices differing by a gauge
transformation might extremize differently, which is what they
observed in their gauge variant minimization procedure.

However, if we take for $S$ the (continuum) Yang-
Mills action
\be
S^{\mbox{{\scriptsize cont}}} = \frac{1}{4g^2}\int_V (F_{\mu \nu}^a)^2
\; , \label{101b}
\ee
we see that the functional derivative in \eqrf{101} gives just
the gauge covariant equation of motion for $A_{\mu}^a(\site)$,
so that the corresponding extremization action $\shat$ is the
gauge {\em invariant\/} quantity
\be
\shat^{\mbox{{\scriptsize cont}}} = \int_V (D_{\mu}
F_{\mu \nu}^a)^2 \; , \label{102}
\ee
disregarding the factor $1/g^2$ in front of
$S^{\mbox{{\scriptsize cont}}}$.
This is not surprising, for the extrema of the action, determined by
$\shat$, should be gauge invariant.
Since the continuum quantity \eqrf{102} is gauge invariant, it must be
possible to define the lattice analogue in a gauge invariant way as well.
It is the purpose of this paper to show how this can be achieved.

In section 2 the gauge invariant extremization action is defined.
It is shown that it has the expected behavior in the limit of lattice
spacing $a\ra 0$
and its structure is visualized in terms of closed Wilson loops.
Section 3 contains a discussion of the
generality of the method and a few comments about
its implementation.
Finally, we briefly discuss physical applications.

\section{Gauge invariant extremization action}

Consider the standard plaquette action for SU(N) lattice gauge theory,
\bea
S &=& \bt \,\sum_{\Box} \lb 1 - \frac{1}{2N} (\mbox{Tr}\;U_{\Box} +
\mbox{Tr}\;U_{\Box}^{\dagger}) \rb
\label{101a} \\
&&\mbox{}\;\;\;\;\;\;\;\;(\bt=2N/g^2)
\nn \\
&=&  - \frac{1}{g^2} \sum_\Box \lb \Tr U_\Box + \Tr U_\Box^\dagger
\rb + \mbox{Const.} \label{101ab}
\eea
We will forget about the irrelevant constant.
The degrees of freedom in this theory are the link variables
$U_\mu(\site)$ which are
matrices in a unitary representation of the gauge group.
Therefore it is appropriate to consider the functional derivatives of the
action \rf{101ab} with respect to $U_\mu(\site)$. Since
the link variable $U_\mu(\site)$
is related to a corresponding continuum gauge field $A_\mu(\site)$ by
the path ordered integral
\thicklines
\setlength{\unitlength}{1.5ex}
\be
U = U_\mu(\site) = {\cal P} \exp \left[-ia\int_0^1 A_\mu(x+ta\hat{\mu})
 \, dt \right] =
\raisebox{-0.25\unitlength}{
\begin{picture}(9,2)
\put(1,1){\circle*{0.4}}
\put(5,1){\circle*{0.4}}
\put(1,1){\vector(1,0){2}}
\put(3,1){\line(1,0){2}}
\put(0.25,-0.5){\mbox{$x$}}
\put(5,-0.5){\mbox{$x+a\hat{\mu}$}}
\end{picture}
}
  \;\;\;,\label{101ac}
\ee
which can be expanded in $a$ to give
\be
U_\mu(\site) = 1 - iaA_\mu(\site) + {\cal O}(a^2) \; , \label{101ad}
\ee
one expects this derivative to become equivalent to the
derivative with respect to
$A_\mu(\site)$ in the limit $a\ra 0$.
In this way, we shall define a manifestly gauge invariant extremization
action on the lattice and show that it reduces to \eqrf{102} to leading
order in the lattice spacing.

We start by rewriting the plaquette action \rf{101ab} in a form that
exhibits the dependence on a particular link $U$ more clearly,
\bea
S &=& - \frac{1}{g^2} \sum_{\stackrel{\rm links}{U}}
\Tr \lb U F_U + U^\dagger F_U^\dagger \rb \label{102z1}\\
&&\mbox{+ terms independent of $U$} \nn
\eea
(from now on we will omit factors of $1/g^2$).
Here $F_U$ is the sum of the $\gm=2(d-1)$ `staples' for the link $U$,
\be
F_{U_\mu(x)} = \sum_\nu \;\;
\lb
\raisebox{-\unitlength}{
\begin{picture}(6,7.5)(0,4)
\put(1,5){\circle*{0.4}}
\put(5,5){\circle*{0.4}}
\put(5,5){\vector(0,1){2}}
\put(5,7){\line(0,1){2}}
\put(5,9){\vector(-1,0){2}}
\put(3,9){\line(-1,0){2}}
\put(1,9){\vector(0,-1){2}}
\put(1,7){\line(0,-1){2}}
\put(0.5,3.5){\mbox{$x$}}
\end{picture}
}
 +
\raisebox{-\unitlength}{
\begin{picture}(6,7.5)(0,4)
\put(1,5){\circle*{0.4}}
\put(5,5){\circle*{0.4}}
\put(5,5){\vector(0,-1){2}}
\put(5,3){\line(0,-1){2}}
\put(5,1){\vector(-1,0){2}}
\put(3,1){\line(-1,0){2}}
\put(1,1){\vector(0,1){2}}
\put(1,3){\line(0,1){2}}
\put(0.5,6){\mbox{$x$}}
\end{picture}
}
\rb \;\;\hspace*{2cm}\;\;
\begin{picture}(6,5)(0,5)
\put(1,5){\vector(1,0){4}}
\put(1,5){\vector(0,1){4}}
\put(4,3.5){\mbox{$\mu$}}
\put(-0.5,8){\mbox{$\nu$}}
\end{picture}
 \; , \label{102z2}
\ee
which upon contraction with $U$ gives the part of the plaquette action
depending on $U$. $\gm$ can be considered as the coordination number for
the interactions.
A staple is defined as
\be
\raisebox{-2\unitlength}{
\begin{picture}(6,6)(0,-1)
\put(1,1){\circle*{0.4}}
\put(5,1){\circle*{0.4}}
\put(5,1){\vector(0,1){2}}
\put(5,3){\line(0,1){2}}
\put(5,5){\vector(-1,0){2}}
\put(3,5){\line(-1,0){2}}
\put(1,5){\vector(0,-1){2}}
\put(1,3){\line(0,-1){2}}
\put(0.5,-0.5){\mbox{$x$}}
\end{picture}
}
= U_\nu(x+a\hat{\mu}) \, U_\mu^\dagger(x+a\hat{\nu}) \, U_\nu^\dagger(x)
\;\;\hspace*{2cm}\;\;
\raisebox{-2\unitlength}{
\begin{picture}(6,6)(0,-1)
\put(1,1){\vector(1,0){4}}
\put(1,1){\vector(0,1){4}}
\put(4,-0.5){\mbox{$\mu$}}
\put(-0.5,4){\mbox{$\nu$}}
\end{picture}
}
\; . \label{102z3}
\ee
$F_U$ is sometimes called the `force'
(hence the notation $F$) for the link $U$. Note that it is not an
element of the SU(N) in general. In the special case of gauge group SU(2) it
can always be written as a constant times an SU(2) matrix.

A variation $\dl U$ in the link $U=U_\mu(\site)$
gives rise to a change in action $\dl S$ equal to
\bea
\dl S &=& - \Tr \left\{ \lb \frac{\dl S}{\dl U} - U^{\dagger}
      \frac{\dl S}{\dl U^{\dagger}} U^{\dagger}\rb \dl U \right\}
      \label{102b} \\
&=& - \Tr \left\{ (UF-(UF)^\dagger)\, \dl U\, U^\dagger \right\}
     \; , \label{102c}
\eea
where we have used that
\be
0 = \dl (U^{\dagger}U) = (\dl U^\dagger)U + U^\dagger \dl U
 \; ,  \label{102f}
\ee
and it is understood that $F=F_U$.
We can visualize $(UF-(UF)^\dagger)$ as
\be
\sum
\;\;\lb
\raisebox{-\unitlength}{
\begin{picture}(6,6)(0,0)
\put(1.5,1){\circle*{0.4}}
\put(1,1.5){\circle*{0.4}}
\put(0.5,-0.5){\mbox{$x$}}
\put(1.5,1){\vector(1,0){2}}
\put(3.5,1){\line(1,0){1.5}}
\put(5,1){\vector(0,1){2}}
\put(5,3){\line(0,1){2}}
\put(5,5){\vector(-1,0){2}}
\put(3,5){\line(-1,0){2}}
\put(1,5){\vector(0,-1){2}}
\put(1,3){\line(0,-1){1.5}}
\end{picture}
}
-
\raisebox{-\unitlength}{
\begin{picture}(6,6)(0,0)
\put(1.5,1){\circle*{0.4}}
\put(1,1.5){\circle*{0.4}}
\put(0.5,-0.5){\mbox{$x$}}
\put(1,1.5){\vector(0,1){2}}
\put(1,3.5){\line(0,1){1.5}}
\put(1,5){\vector(1,0){2}}
\put(3,5){\line(1,0){2}}
\put(5,5){\vector(0,-1){2}}
\put(5,3){\line(0,-1){2}}
\put(5,1){\vector(-1,0){2}}
\put(3,1){\line(-1,0){1.5}}
\end{picture}
}
\rb
\label{102ga}
\ee
where the sum runs over the $\gm=2(d-1)$ staple directions.

We now define the gauge invariant lattice analogue of the
extremization action $\shat$ by
\bea
\shat &=& \sum_{\stackrel{\rm links}{U}}\half \Tr \left\{
(UF - (UF)^\dagger)^\dagger (UF-(UF)^\dagger)
\right\}  \label{104} \\
&=& \sum_{\stackrel{\rm links}{U}} \half \Tr \{
2F^\dagger F - F U F U - (F U F U)^\dagger \}
 \; . \label{105}
\eea
This is a sum over closed Wilson lines of lengths 6 and 8,
arising from contractions of the diagrams in \eqrf{102ga}.
Obviously, $\shat$ is gauge invariant.

Before giving a more detailed survey of the diagrams
we shall show that $\shat$ reduces to its continuum
counterpart \rf{102} in the limit $a\ra 0$.
For simplicity, we shall do so for the abelian case. The non-abelian
generalization will follow easily.
For completeness we mention that for the gauge group U(1) the plaquette
action is
\be
S_{U(1)} = - \frac{1}{2g^2} \, \sum_\Box \, (\Box + \Box^*)
\label{107}
\ee
and the appropriate definition of $\shat$ is
\be
\shat_{U(1)} = \qrt \sum_{\stackrel{\rm links}{U}}
(UF - (UF)^*)^* (UF-(UF)^*)
  \; . \label{109}
\ee

Consider the expression for $UF - (UF)^*$ in \eqrf{102ga}
with $U=U_\mu(\site)$.
For each $\nu\neq\mu$ the sum contains a contribution in the positive and
one in the negative $\nu$-direction.
The term from the positive $\nu$-direction is calculated as follows. First
one computes the expansion of the plaquette around its centre,
\be
\raisebox{-\unitlength}{
\begin{picture}(6,6)(0,0)
\put(1.5,1){\circle*{0.4}}
\put(1,1.5){\circle*{0.4}}
\put(0.5,-0.5){\mbox{$x$}}
\put(1.5,1){\vector(1,0){2}}
\put(3.5,1){\line(1,0){1.5}}
\put(5,1){\vector(0,1){2}}
\put(5,3){\line(0,1){2}}
\put(5,5){\vector(-1,0){2}}
\put(3,5){\line(-1,0){2}}
\put(1,5){\vector(0,-1){2}}
\put(1,3){\line(0,-1){1.5}}
\end{picture}
}
=
\exp \, [-ia^2 F_{\mu\nu}(x+\hhalf a \hat{\mu} + \hhalf a \hat{\nu}) +
\calo(a^4)]
\; , \label{210}
\ee
so that
\be
\lb
\raisebox{-\unitlength}{
\begin{picture}(6,6)(0,0)
\put(1.5,1){\circle*{0.4}}
\put(1,1.5){\circle*{0.4}}
\put(0.5,-0.5){\mbox{$x$}}
\put(1.5,1){\vector(1,0){2}}
\put(3.5,1){\line(1,0){1.5}}
\put(5,1){\vector(0,1){2}}
\put(5,3){\line(0,1){2}}
\put(5,5){\vector(-1,0){2}}
\put(3,5){\line(-1,0){2}}
\put(1,5){\vector(0,-1){2}}
\put(1,3){\line(0,-1){1.5}}
\end{picture}
}
-
\raisebox{-\unitlength}{
\begin{picture}(6,6)(0,0)
\put(1.5,1){\circle*{0.4}}
\put(1,1.5){\circle*{0.4}}
\put(0.5,-0.5){\mbox{$x$}}
\put(1,1.5){\vector(0,1){2}}
\put(1,3.5){\line(0,1){1.5}}
\put(1,5){\vector(1,0){2}}
\put(3,5){\line(1,0){2}}
\put(5,5){\vector(0,-1){2}}
\put(5,3){\line(0,-1){2}}
\put(5,1){\vector(-1,0){2}}
\put(3,1){\line(-1,0){1.5}}
\end{picture}
}
\rb
=
-2i \sin \, [a^2 F_{\mu\nu}(x+\hhalf a \hat{\mu} + \hhalf a \hat{\nu})
+ \calo(a^4)] \; . \label{211}
\ee
If the expansion in \eqrf{210} were not done around the centre of the
plaquette, $\calo(a^3)$ terms would be present as well.
The contribution from the negative $\nu$-direction is
\be
+2i \sin \, [a^2 F_{\mu\nu}(x+\hhalf a \hat{\mu} - \hhalf a \hat{\nu})
+ \calo(a^4)]  \label{212}
\ee
and the sum of eqs.~\rf{211} and \rf{212} can be expanded around $x+\half a
\hat{\mu}$ and subsequently summed over $\nu$ to give
for $UF - (UF)^*$ in \eqrf{102ga}:
\be
\sum
\;\;\lb
\raisebox{-\unitlength}{
\begin{picture}(6,6)(0,0)
\put(1.5,1){\circle*{0.4}}
\put(1,1.5){\circle*{0.4}}
\put(0.5,-0.5){\mbox{$x$}}
\put(1.5,1){\vector(1,0){2}}
\put(3.5,1){\line(1,0){1.5}}
\put(5,1){\vector(0,1){2}}
\put(5,3){\line(0,1){2}}
\put(5,5){\vector(-1,0){2}}
\put(3,5){\line(-1,0){2}}
\put(1,5){\vector(0,-1){2}}
\put(1,3){\line(0,-1){1.5}}
\end{picture}
}
-
\raisebox{-\unitlength}{
\begin{picture}(6,6)(0,0)
\put(1.5,1){\circle*{0.4}}
\put(1,1.5){\circle*{0.4}}
\put(0.5,-0.5){\mbox{$x$}}
\put(1,1.5){\vector(0,1){2}}
\put(1,3.5){\line(0,1){1.5}}
\put(1,5){\vector(1,0){2}}
\put(3,5){\line(1,0){2}}
\put(5,5){\vector(0,-1){2}}
\put(5,3){\line(0,-1){2}}
\put(5,1){\vector(-1,0){2}}
\put(3,1){\line(-1,0){1.5}}
\end{picture}
}
\rb
=
-2ia^3 \sum_\nu \partial_\nu F_{\mu\nu}(x+\hhalf a \hat{\mu})
+ \calo(a^5)  \; . \label{213}
\ee
Terms of $\calo(a^4)$ cancel out.
Finally, this result is inserted in \eqrf{109}, giving
\be
\shat_{U(1)} = \int_V a^2 (\partial_\nu F_{\mu\nu} )^2 +\calo(a^4)
\label{214}
\ee
which is the abelian version of \eqrf{102} to leading order in $a$,
as promised.
The extra factor of $a^2$ here is explained by \eqrf{101ad} and serves
to make $\shat$ dimensionless.
This result can be generalized to the non-abelian case immediately, by
replacing the derivatives by covariant ones and adjusting the
coefficients appropriately.

Now we turn to an overview of the diagrams occurring in $\shat$
(\ref{104}--\ref{105}).
Each link $U$ in the summation \eqrf{105}
yields $(2\gm)^2$ closed Wilson loops,
with the appropriate signs (recall that $\gm=2(d-1)$).
In 2, 3 and 4 dimensions, $4\gm^2= 16, 64, 144$ respectively.
These huge numbers include a number of trivial
and double-counted diagrams as well as hermitian conjugates, however.
\setlength{\unitlength}{3ex}
\begin{figure}
\begin{picture}(30,12)
\put(0,2){%
\raisebox{-\unitlength}{
\begin{picture}(6,7.5)(0,4)
\put(5,5){\vector(0,1){2}}
\put(5,7){\line(0,1){2}}
\put(5,9){\vector(-1,0){2}}
\put(3,9){\line(-1,0){2}}
\put(1,9){\vector(0,-1){2}}
\put(1,7){\line(0,-1){2}}
\put(1,5){\vector(3,4){0.8}}
\put(1.8,6.066667){\line(3,4){0.8}}
\put(2.6,7.133333){\vector(1,0){2}}
\put(4.6,7.133333){\line(1,0){2}}
\put(6.6,7.133333){\vector(-3,-4){0.8}}
\put(5.8,6.066667){\line(-3,-4){0.8}}
\put(1,5){\dashbox{0.2}(4,0){}}
\put(3,3.5){\mbox{$a$}}
\end{picture}
}
}
\put(7,2){%
\raisebox{-\unitlength}{
\begin{picture}(6,7.5)(0,4)
\put(5,5){\vector(0,1){2}}
\put(5,7){\line(0,1){2}}
\put(5,9){\vector(-1,0){2}}
\put(3,9){\line(-1,0){2}}
\put(1,9){\vector(3,4){0.8}}
\put(1.8,10.066667){\line(3,4){0.8}}
\put(2.6,11.133333){\vector(0,-1){2}}
\put(2.6,9.133333){\line(0,-1){2}}
\put(2.6,7.133333){\vector(1,0){2}}
\put(4.6,7.133333){\line(1,0){2}}
\put(6.6,7.133333){\vector(-3,-4){0.8}}
\put(5.8,6.066667){\line(-3,-4){0.8}}
\put(3,3.5){\mbox{$b$}}
\end{picture}
}
}
\put(14,2){%
\raisebox{-\unitlength}{
\begin{picture}(6,7.5)(0,4)
\put(4.7,5){\vector(0,1){2}}
\put(4.7,7){\line(0,1){1.6}}
\put(4.7,8.6){\vector(-1,0){2}}
\put(2.7,8.6){\line(-1,0){2}}
\put(0.7,8.6){\vector(0,-1){2}}
\put(0.7,6.6){\line(0,-1){2}}
\put(0.7,4.6){\vector(1,0){2}}
\put(2.7,4.6){\line(1,0){2}}
\put(4.7,4.6){\vector(3,4){1.1}}
\put(5.8,6.066667){\line(3,4){0.8}}
\put(6.6,7.133333){\vector(-1,0){2.5}}
\put(4.1,7.133333){\line(-1,0){1.5}}
\put(2.6,7.133333){\vector(-3,-4){0.8}}
\put(1.8,6.066667){\line(-3,-4){0.8}}
\put(1,5){\vector(1,0){2}}
\put(3,5){\line(1,0){1.7}}
\put(3,3.5){\mbox{$c$}}
\end{picture}
}
}
\put(21,2){%
\raisebox{-\unitlength}{
\begin{picture}(6,7.5)(0,4)
\put(5,4.6){\vector(0,1){2}}
\put(5,6.6){\line(0,1){2.4}}
\put(5,9){\vector(-1,0){2}}
\put(3,9){\line(-1,0){2}}
\put(1,9){\vector(0,-1){2}}
\put(1,7){\line(0,-1){2}}
\put(1,5){\vector(1,0){2}}
\put(3,5){\line(1,0){1.7}}
\put(4.7,5){\vector(0,1){2}}
\put(4.7,7){\line(0,1){1.6}}
\put(4.7,8.6){\vector(-1,0){2}}
\put(2.7,8.6){\line(-1,0){2}}
\put(0.7,8.6){\vector(0,-1){2}}
\put(0.7,6.6){\line(0,-1){2}}
\put(0.7,4.6){\vector(1,0){2}}
\put(2.7,4.6){\line(1,0){2.3}}
\put(3,3.5){\mbox{$d$}}
\end{picture}
}
}
\end{picture}

\caption{
$\mbox{}$\protect\\
$a$. Length 6 Wilson loop contributing to $\shat$.\protect\\
$b$. Non-contributing length 6 diagram.\protect\\
$c$. Contributing diagram of length 8.\protect\\
$d$. `Double plaquette'.}
\end{figure}

The first term in \eqrf{105} contributes $\gm^2$ Wilson loops of length 6 in
lattice units, each counted twice.
These are the diagrams
`hinging' around the link $U_{\mu}(\site)$, but independent of it,
together with their adjoints, see fig.~1$a$.
Of these, $\gm$ give the unit matrix, another $\gm$ are the planar
$2\times 1$ loops, and the remaining $\gm(\gm-2)$ do not lie in a plane.
Note that the symmetric length 6 Wilson loops winding
around a cube as depicted in fig.~1$b$ are excluded.
The other terms in \eqrf{105} give the length 8 Wilson loops
containing the link $U$ twice (fig.~1$c$) and their conjugates.
These $2\gm^2$ diagrams include
the $\gm$ `double plaquettes' shown in fig.~1$d$ with their conjugates.
In these double plaquette diagrams, the loop winds around a plaquette
twice before closing on itself.

Note how nicely this
method provides one with a lattice action reducing to the
continuum derivative action \eqrf{102} in lowest order in the lattice
spacing $a$.
It is also interesting to observe that,
apart from being gauge invariant, \eqrf{102}
preserves Lorentz-invariance.
This implies that $\shat$ cannot be used as an improvement term for the
plaquette action. The
${\cal O}(a^6)$ term in the expansion of the plaquette action which is
canceled in tree-level improved actions \cite{improved} by adding
$2\times 1$ rectangular Wilson loops, are of the form
\be
\sum_{\mu,\nu} D_\nu F_{\nu\mu} \, D_\nu F_{\nu\mu} \nn
\ee
which does not equal $\sum_\mu (\sum_\nu D_\nu F_{\nu\mu})^2$.

\section{Further remarks}

It should be clear that the applicability of the
procedure described in this
paper is not restricted to the standard plaquette action.
For any lattice action $S$, written as a sum over closed Wilson
lines, the quantity
\be
\shat = \half \sum_U \Tr \left|\frac{\dl S}{\dl U} \right|^2 \label{301}
\ee
provides a gauge invariant extremization action.
The calculational recipe of $\shat$ can be visualized as follows.
Cut the link $U$ out of the $U$-dependent closed
loops occurring in $S$, and paste the resulting open-ended lines
together (with the correct signs) to form closed loops of $\shat$.

Furthermore, the extremization action \eqrf{301} can be defined
for any gauge group, including discrete groups such as $\Z_N$,
the link variables being group elements.
Of course, it is not possible to define a continuum limit
of the extremization action (nor the action itself) if the
gauge group is discrete.
This does not limit the use of the extremization
action for lattice gauge theories with discrete groups, however.

We should also make some remarks about implementation of the
suggested extremization procedure.
First of all, in four dimensions the number of diagrams contributing
to $\shat$ for each link is large so that calculation
and minimization of $\shat$ can be computationally quite expensive.
In two and three dimensions this problem is less severe and the method
may find its applications.
Furthermore, minimization of the extremization action $\shat$
is not straightforward.
The dependence of $\shat$ on a particular link $U$ is of the non-linear
form
\be
\Tr [MUMU + NU + \mbox{hermitian conjugate}] \; , \label{302}
\ee
where both $M$ and $N$ are sums over $SU(N)$ matrices (or U(1) numbers).
An iterative procedure is probably
required to carry out the local minimization step.
Finally, one may need special algorithms to speed up the long
distance modes, as in ref.~\cite{dunmaw1}.

We end with a few remarks concerning the application of
extremization to physical problems, in the light of the work of
Duncan and Mawhinney \cite{dunmaw1}.
It would be interesting to see if
their results are confirmed when the gauge invariant procedure proposed
here is used to extremize configurations.
In any case, gauge dependence in the results should
be absent when the gauge invariant procedure is used.

In the simulation of
three-dimensional SU(2) gauge theory in ref.~\cite{dunmaw1} it was
found that during extremization the action itself decreased as well.
At first sight, this is somewhat surprising, as it is hard to see why a
quantity should decrease when its derivatives are minimized.
One wonders whether the fact that the functional derivatives
in the gauge variant extremization procedure are not exact
lattice derivatives has any influence here.
Similarly, it is surprising that a strong correlation was noticed
between lumps in the action density after cooling a configuration and
after extremizing the same configuration.
Finally, it was observed
that the string tension disappeared after continued extremization.
These are all interesting problems that merit further investigation.

\ACK
It is a pleasure to thank M.~Teper
for helpful comments on the manuscript.


\small
\begin{thebibliography}{99}
\bibitem{cooling} B.~Berg, \plb{104} (1981) 475; \\
M.~L\"uscher, \npb{200} (1982) 61; \\
J.~Hoek, M.~Teper and J.~Waterhouse, \npb{288} (1987) 589
\bibitem{dunmaw1} A.~Duncan and R.D.~Mawhinney, \nproc{26} (1992)
444; \\ \plb{282} (1992) 423
\bibitem{improved} K.~Symanzik, \npb{226} (1983) 187, 205
\end{thebibliography}
\end{document}